\documentclass[a4paper,11pt]{article}
\usepackage{pos}

\usepackage{placeins}

\title{Recent quarkonium measurements in small systems with the ALICE detector at the LHC}
\ShortTitle{Recent quarkonium measurments in small systems with the ALICE detector}

\author*[a]{Manuel Guittière}
\author[]{on behalf of the ALICE Collaboration}

\affiliation[a]{Laboratoire de physique subatomique et des technologies associées - Subatech, CNRS\\
  Nantes, France}

\emailAdd{manuel.guittiere@cern.ch}

\abstract{
At the LHC collision energies, multiple parton interactions (MPI) are a key ingredient for particle production models including
hard scale processes like heavy-quark production (charm and beauty).
Quarkonium measurements in high-multiplicity proton-proton (pp) collisions can shed light on the role
of MPI at such hard momentum scales, as well as on the interplay between hard and soft particle production mechanisms.
In addition, quarkonium production measurements in minimum bias pp collisions, besides serving as a reference
for heavy-ion collisions collected at the same center-of-mass energy, represent a benchmark test of various QCD based models.
\newline In this contribution, the latest quarkonium measurements performed by the ALICE collaboration in pp
collisions at several center-of-mass energies are presented. A comprehensive study of the multiplicity dependence of the quarkonium production at
$\sqrt{s} = 13$~TeV, based on minimum bias and high-multiplicity triggered events, is also presented.
Such measurements include $\psi$(2S) production at forward rapidity as a function of the charged particle
multiplicity density, as well as the latest multiplicity dependent inclusive J/$\psi$ production measurements
at midrapidity, based on multiplicity estimators covering different pseudorapidity regions.
Similar multiplicity dependent measurements in p--Pb collisions at center-of-mass energies of
$\sqrt{s_{\rm NN}} =$ 5.02 and 8.16 TeV are also presented. The reported results are compared with available theoretical
model calculations.
}

\FullConference{%
  HardProbes2020\\
  1-6 June 2020\\
  Austin, Texas}

\begin{document}
\maketitle

\section{Introduction}

Quarkonia production represents an interesting observable to probe hard processes occurring in different hadronic collision systems and to study
the effects of high multiplicity environments created in such high-energy collisions at the LHC. Recent cross section measurements performed in proton-proton (pp)
collisions by ALICE provide new measurements of the nuclear modification factor in proton-lead (p--Pb) and lead-lead (Pb--Pb) collision systems at the same energy.
\newline
\indent Measurements of multiplicity dependent quarkonium production in small systems are performed to study the effect of multi-parton interactions (MPI) and the interplay between
hard and soft particle production at the LHC collision energies. The multiplicity dependent production of upsilon excited states ($\Upsilon$(2S) and $\Upsilon$(3S)) with respect to the ground state ($\Upsilon$(1S))
has been recently measured by the CMS Collaboration \cite{CMSUpsilonInvestigation} in high multiplicity pp collisions. In addition, recently published ALICE results in p--Pb \cite{ALICERpPb8TeV}
reported a stronger suppression of the $\psi$(2S) compared to the $\rm{J}/\psi$ ground state at backward rapidity ($-4.46 < y_{\rm cms} < -2.96$), well reproduced by the comovers model \cite{FerreiroComovers}.
Motivated by these observations, the recent multiplicity dependent measurements reported in this contribution aim to improve the understanding of the non-trivial underlying
mechanisms in high-energy pp and p--Pb collisions at the LHC.

\section{Experimental setup and analysis strategy}

The quarkonium measurements reported in this contribution were performed in different rapidity regions of the ALICE detector \cite{ALICEDetector}.
At central rapidity ($|y| < 0.9$), $\rm{J}/\psi$ are measured in their dielectron decay channel, relying mainly on the combined tracking capabilities of the Inner Tracking System (ITS) and the Time Projection Chamber (TPC).
The muon spectrometer located at forward rapidity ($2.5 < y < 4$) was used to measure charmonium and bottomonium in their dimuon decay channel. In asymmetric p--Pb collisions the center of mass of the system is shifted resulting in an acceptance
of $2.03 < y_{\rm cms} < 3.53$ at forward rapidity (p-going direction) and $-4.46 < y_{\rm cms} < -2.96$ at backward rapidity (Pb-going direction) for the muon spectrometer.
\newline
\indent Multiplicity measurements are also performed in two rapidity regions. The V0 detector, composed of two scintillator arrays located at forward ($-3.7 < \eta < -1.7$) and backward ($2.8 < \eta < 5.1$) pseudorapidity, provides
the charged particle multiplicity estimation at large rapidity and the Silicon Pixel Detector (SPD: two innermost layers of the ITS) plays this role at midrapidity ($|y| < 1$).
Here, the number of charged particle tracks is equalized along the beam axis ($z$) in order to correct the tracking acceptance or efficiency of the SPD. Then, the
multiplicity, defined as the charged particle pseudorapidity density (d$N_{\rm ch}/$d$\eta$), is measured using simulated multiplicity distributions based on PYTHIA-8 \cite{PYTHIA82} and EPOS-LHC \cite{EPOSLHC} event generators.

\section{Measurements of production cross sections}

\begin{figure}[!ht]
  \centering
  \includegraphics[scale=0.37]{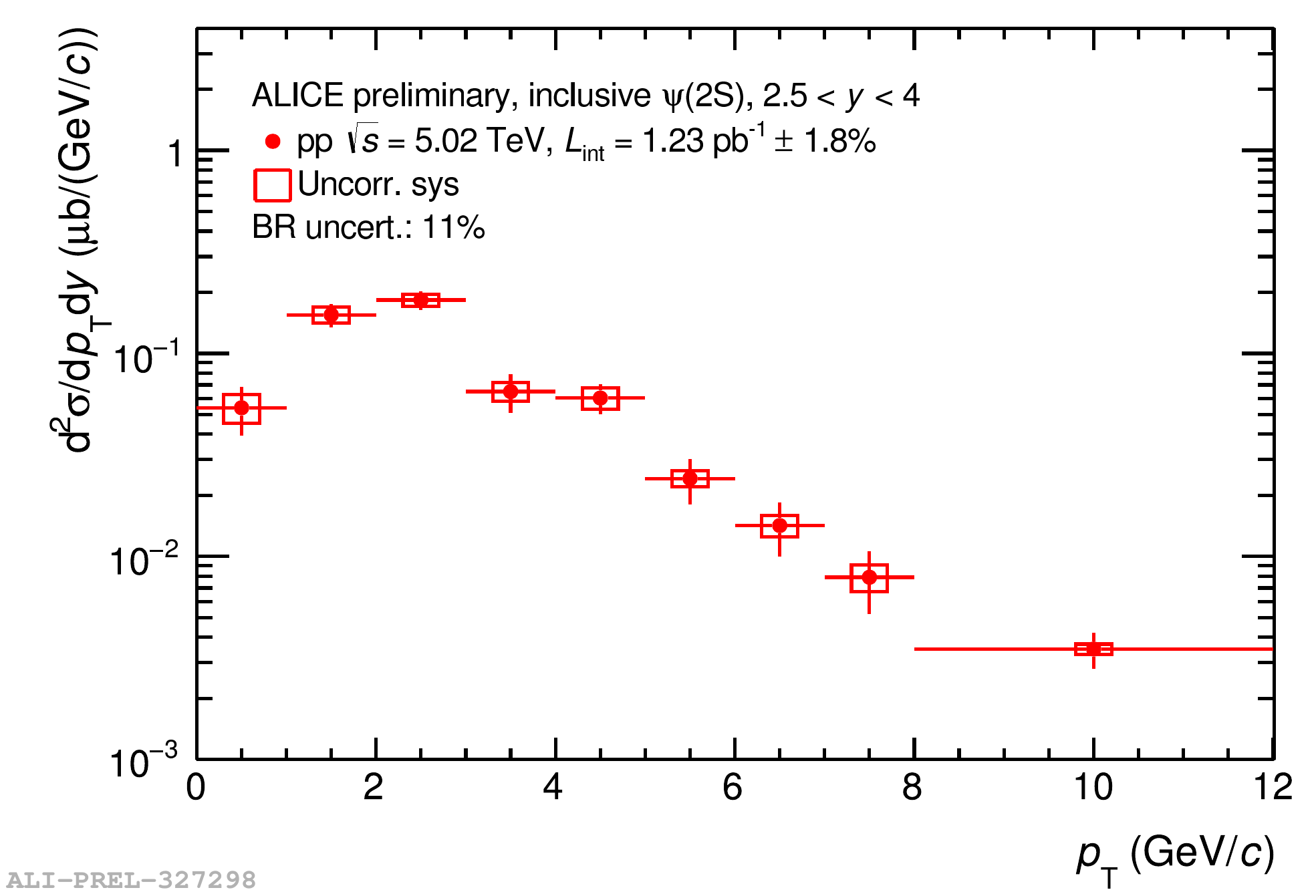}
  \includegraphics[scale=0.353]{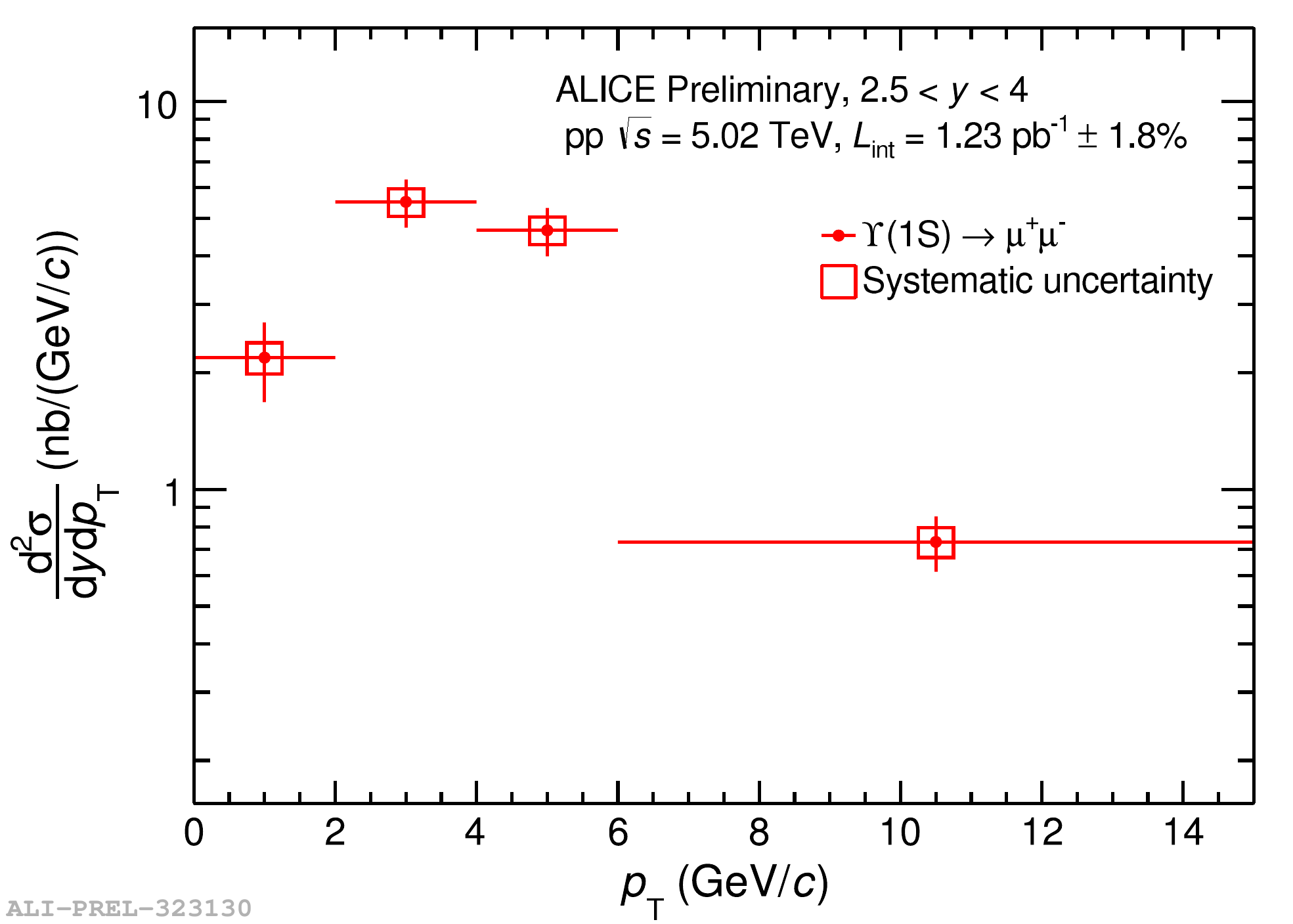}
  \caption{Inclusive $p_{\rm T}$ differential $\psi$(2S) (left) and $\Upsilon$(1S) (right) cross section in pp collisions at $\sqrt{s} = 5.02$~TeV.}
  \label{Psi2SUps1S5TeVCrossSec}
\end{figure}

The ALICE Collaboration has recently released several inclusive quarkonium cross section measurements in pp collisions at $\sqrt{s} = 5.02$~TeV in both charm and beauty sectors.
The $\rm{J}/\psi$ cross section was measured up to $p_{\rm T}$ = 10~GeV/$c$ at midrapidity and up to 20~GeV/$c$ at forward rapidity. In reference \cite{JPsiCrossSec}, the experimental results are compared with non-relativistic QCD (NRQCD) calculations \cite{NLONRQCD1,NLONRQCD2}
coupled with the fixed order + next to leading logarithm (FONLL) approach \cite{FONLL} for the non-prompt $\rm{J}/\psi$ and with color glass condensate (CGC) calculations \cite{NRQCDCGC} for the prompt component.
Such models are able to reproduce the data on the whole $p_{\rm T}$ range.
\newline
\indent Figure \ref{Psi2SUps1S5TeVCrossSec} (left) shows the first $p_{\rm T}$ differential measurement of the $\psi$(2S) cross section at $\sqrt{s} = 5.02$~TeV. The corresponding rapidity differential cross section was also measured.
This new measurement is a crucial ingredient for future $\psi$(2S) nuclear modification factor ($R_{\rm AA}$) measurements in Pb--Pb at the same energy.

\indent The differential $\psi$(2S) measurements were made possible thanks to the large data sample collected by ALICE in 2017. This is also reflected in the beauty sector where the $p_{\rm T}$ (Fig. \ref{Psi2SUps1S5TeVCrossSec} (right)) and rapidity differential $\Upsilon$(1S) cross sections were also measured for the first time by ALICE at $\sqrt{s} = 5.02$~TeV. Integrated cross sections measurements of $\Upsilon$(2S) and $\Upsilon$(3S) were also performed.
As for the $\psi$(2S) resonance, these results will allow for $\Upsilon$(nS) $R_{\rm AA}$ measurements at this energy.

\section{Multiplicity dependent measurements}

\begin{figure}[!ht]
  \centering
  \includegraphics[scale=0.33]{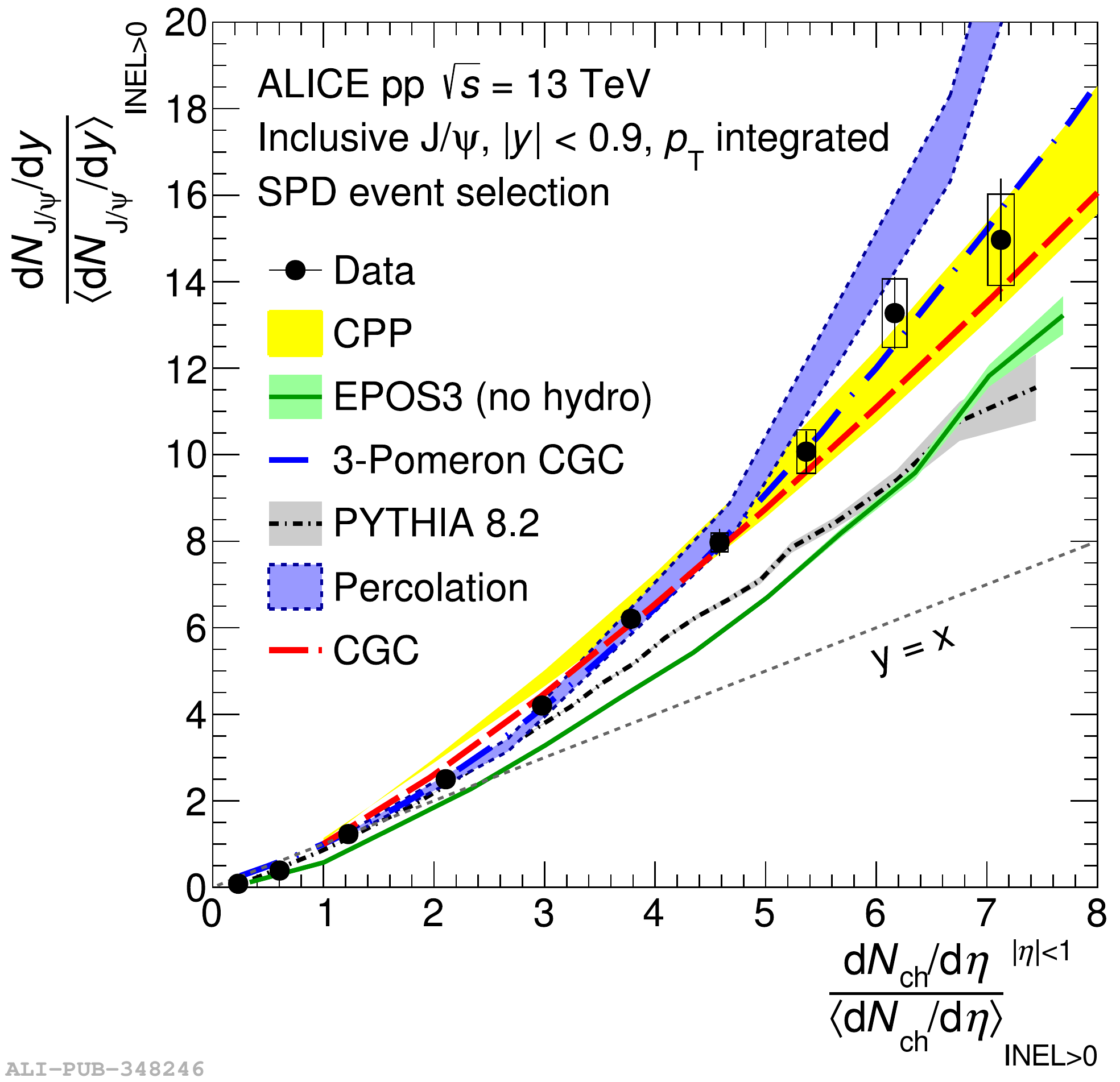}
  \includegraphics[scale=0.33]{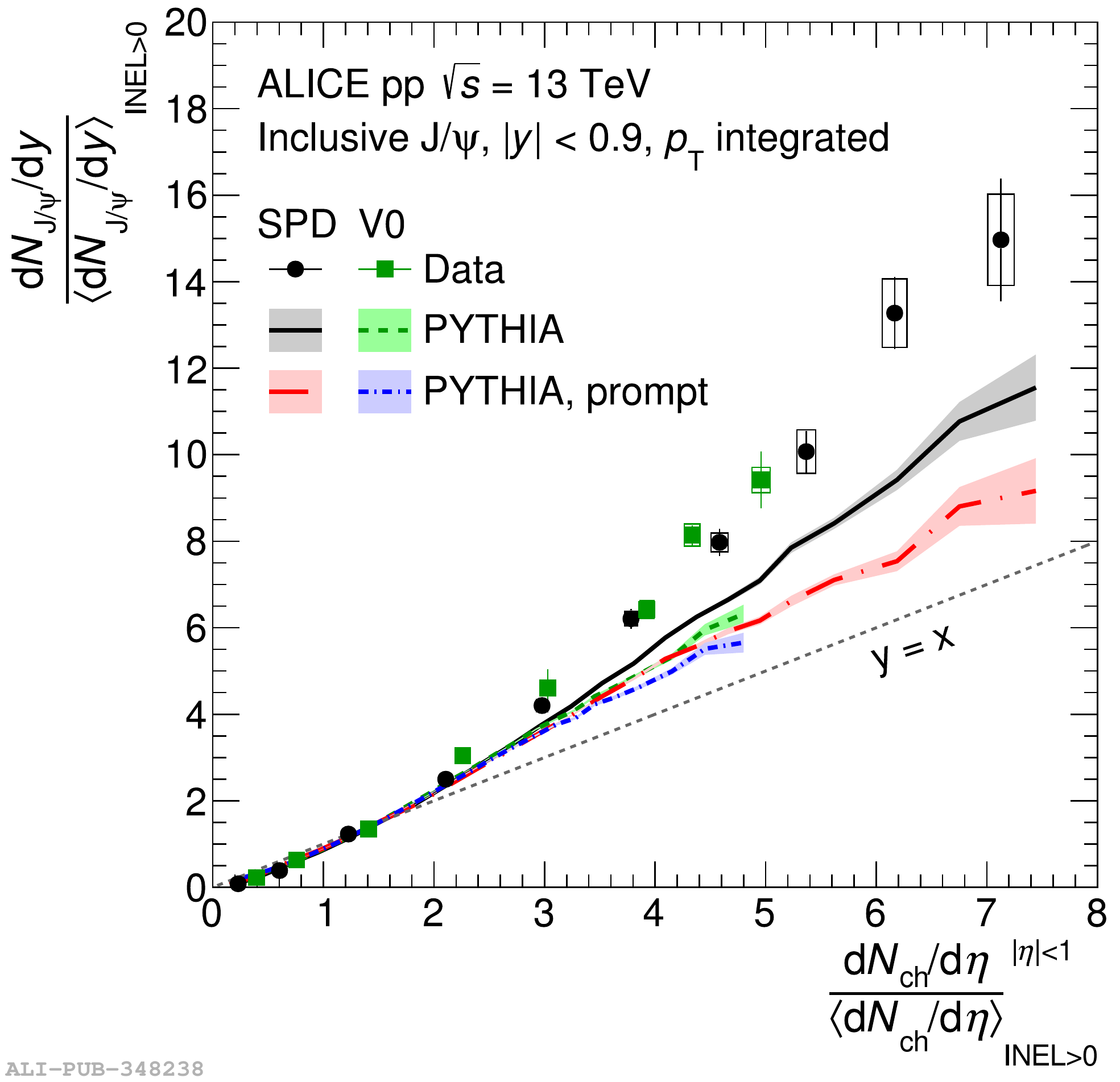}
  \caption{$\rm{J}/\psi$ relative yield at midrapidity as a function of relative charged particle multiplicity at midrapidity (SPD) and forward rapidity (V0) in pp collisions at $\sqrt{s} = 13$~TeV compared to various theoretical models.}
  \label{JPsiMid13TeV}
\end{figure}

Figure \ref{JPsiMid13TeV} shows the recently published measurement of the multiplicity dependent $\rm{J}/\psi$ relative yield at midrapidity in pp collisions at $\sqrt{s} = 13$~TeV \cite{JPsiMidVsMult13TeV}. The multiplicity was measured at both mid and forward rapidity showing no significant difference in the stronger than linear multiplicity dependence of the $\rm{J}/\psi$ relative yield.
The results are compared with various models (left panel of Fig. \ref{JPsiMid13TeV}) that predict such behavior with increasing multiplicity. Among them, the CGC \cite{CGCRef}, the coherent particle production (CPP) \cite{CPPRef}, and the 3-Pomeron CGC \cite{3PomeronCGCRef} models show better agreement with the data. Different mechanisms included in the models are responsible for the predicted stronger than linear production of $\rm{J}/\psi$ such as the
Color String Reconnection (CSR), percolation and gluon saturation. It is worth noticing that a significant reduction of the correlation is observed in predictions from PYTHIA when only including the prompt component (right panel of Fig. \ref{JPsiMid13TeV}).

\indent The multiplicity dependence of $\rm{J}/\psi$ and $\psi$(2S) relative yields was also measured at forward rapidity in pp collisions. Results at $\sqrt{s} = 5.02$~TeV for $\rm{J}/\psi$ and at $\sqrt{s} = 13$~TeV for both charmonia show a linear dependence on relative multiplicity (measured at midrapidity) unlike the $\rm{J}/\psi$ measurement at midrapidity.
Similar measurements were also performed on the $\Upsilon$(1S) and $\Upsilon$(2S) resonances at $\sqrt{s} = 13$~TeV showing a linear behavior compatible with the charmonium results.

\begin{figure}[!ht]
  \centering
  \includegraphics[scale=0.375]{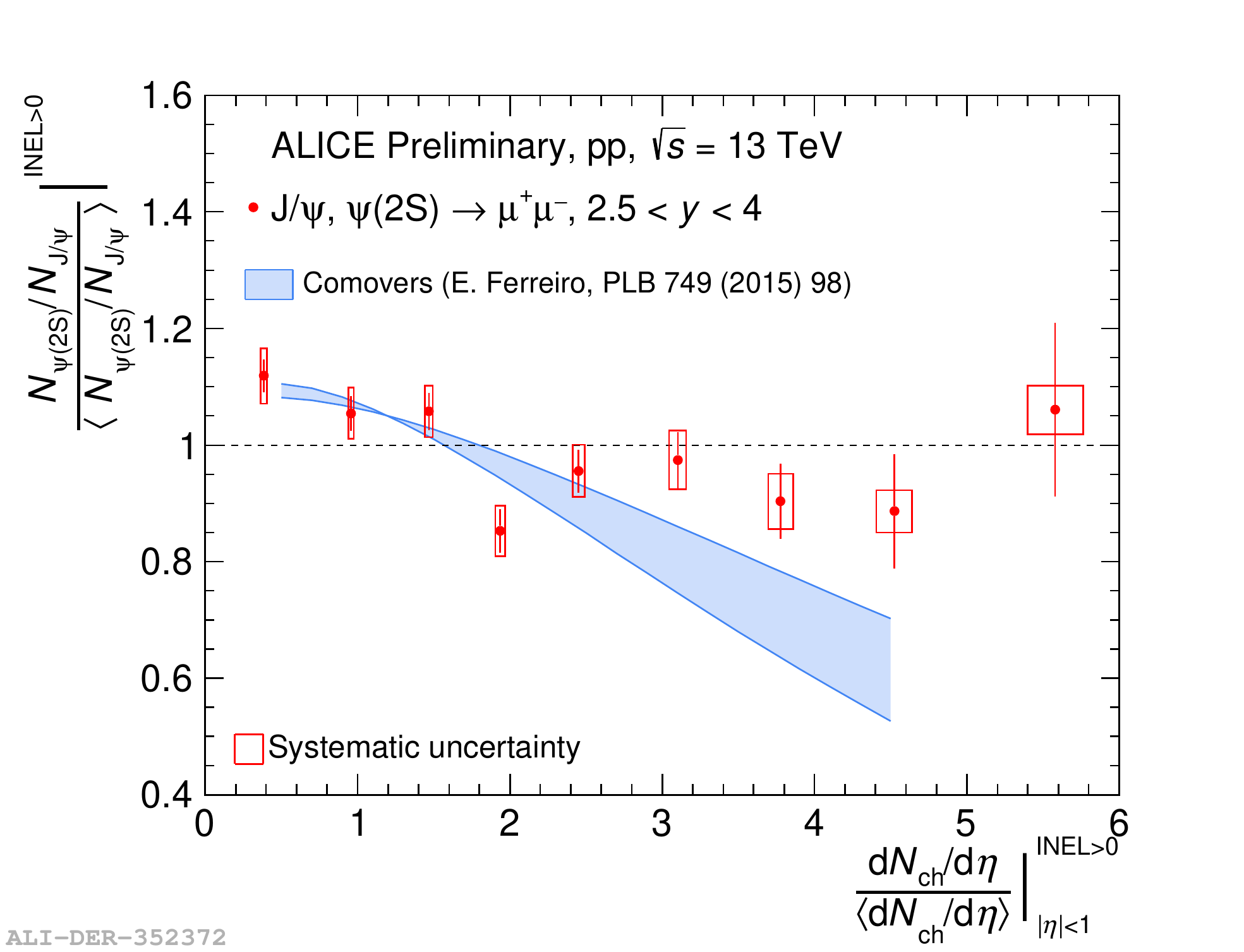}
  \includegraphics[scale=0.375]{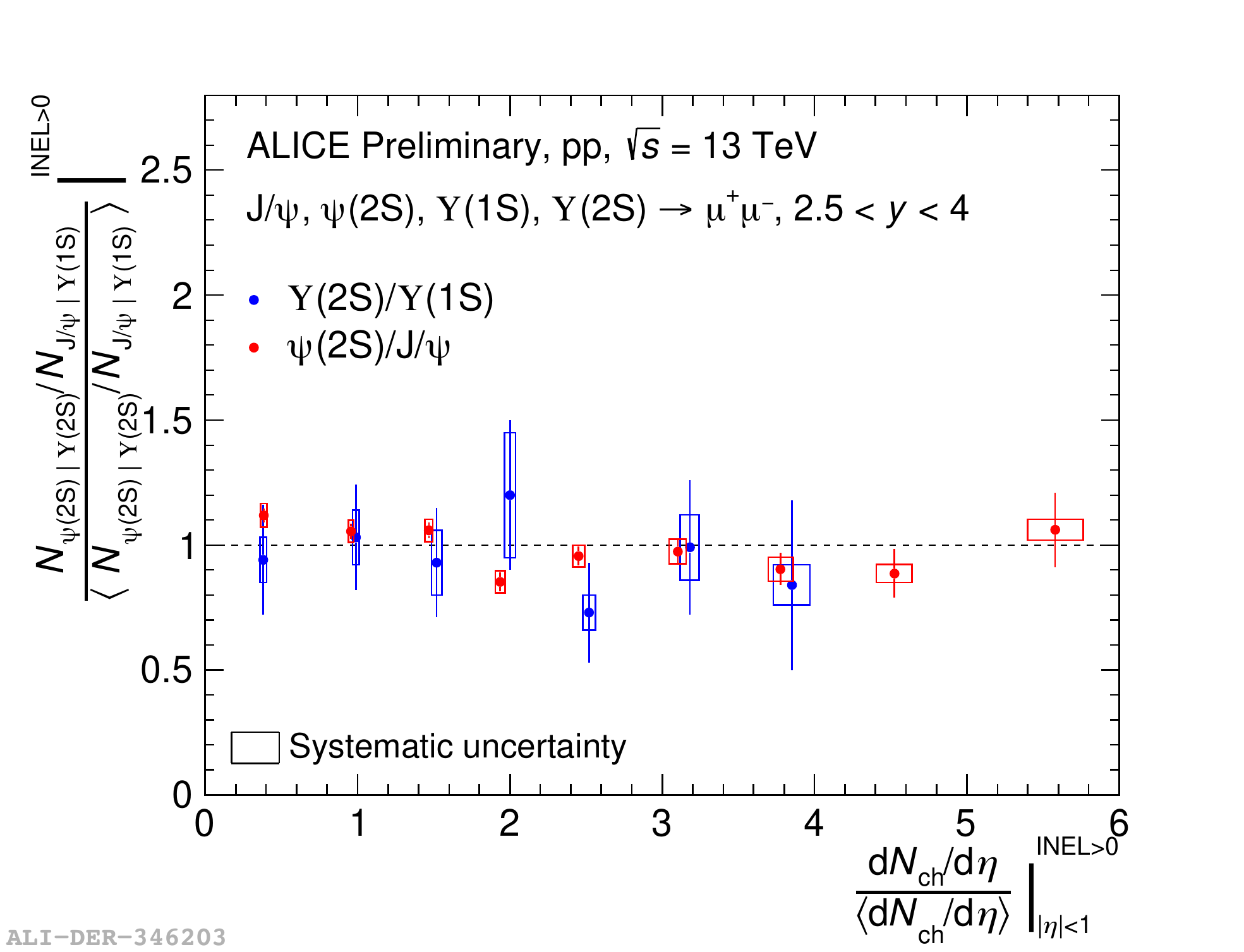}
  \caption{$\psi$(2S)/$\rm{J}/\psi$ relative yield at forward rapidity as a function of relative charged particle multiplicity at midrapidity in pp collisions at $\sqrt{s} = 13$~TeV compared to predictions based on the comovers model (left) and to the corresponding $\Upsilon$(2S)/$\Upsilon$(1S) relative yield (right).}
  \label{2SOver1S}
\end{figure}

\indent Figure \ref{2SOver1S} shows the self-normalized $\psi$(2S) to $\rm{J}/\psi$ relative yield compared to predictions from the comovers model \cite{FerreiroComovers} (left) and to the $\Upsilon$(2S) to $\Upsilon$(1S) ratio as a function of the charged particle multiplicity (right).
Considering the sizeable systematic uncertainties, dominated by the one on the $\psi$(2S) signal extraction, it is not possible to conclude on a significant multiplicity dependence (maximum deviation from unity: 2.2$\sigma$ in the first bin).
It is worth noting that a linear fit performed on these results (negative slope $\approx -0.045$) is in better agreement with the data than a constant fit. The calculations from the comovers model \cite{FerreiroComovers} slightly overpredict a potential $\psi$(2S) suppression with respect to $\rm{J}/\psi$ at high multiplicity.
The decorrelation between the rapidity of the multiplicity and the quarkonium measurements is not taken into account in these calculations. This rapidity gap might partially explain the difference observed between the comovers predictions and the data.
Within the uncertainties, the charmonium and bottomonium results are compatible.
\newline
\indent The multiplicity dependent $\rm{J}/\psi$ production was measured in p--Pb collisions at $\sqrt{s} = 8.16$~TeV and recently published in \cite{JPsipPbVsMult8TeV}. At backward rapidity, the $\rm{J}/\psi$ production shows a slightly stronger than linear dependence on multiplicity. However, at forward rapidity, a significant weaker than linear dependence is observed highlighting the effect of the presence of a nucleus in the initial state with respect to pp results.

\section{Summary and conclusion}

In this contribution, new cross section measurements of charmonium and bottomonium in pp collisions at $\sqrt{s} = 5.02$~TeV were presented. They will allow for new measurements of the nuclear modification factor in p--Pb and Pb--Pb at the same energy. Multiplicity dependent quarkonium measurements in pp collisions at $\sqrt{s} = 13$~TeV as well as $\rm{J}/\psi$ results in p--Pb at $\sqrt{s} = 8.16$~TeV were also reported.
A different behavior was observed between the measurements in different rapidity regions. midrapidity results are well described by different models. However, further investigations and the increasing LHC Run 3 precision will be necessary to improve the understanding of the interplay between hard and soft processes in high multiplicity pp collisions.

\end{document}